\documentstyle[12pt]{article}
\newcommand{\ver}{U^{r}(z)}

\newcommand{\ves}{U^{s}(\xi)}

\newcommand{\1}{\begin{equation}Q^{\mu}(z)=q^{\mu}-ip^{\mu}lnz+i\sum_{n\neq 0}
\frac{\alpha^{\mu}_{n}}{n}z^{-n}\end{equation}}

\newcommand{\gver}{U^{\{r,r_{i}\}}_{\{(n_{i})\}}(z)}

\newcommand{\gves}{U^{\{s,s_{j}\}}_{\{(n_{j})\}}(\xi)}

\newcommand{\mgcef}
{{\cal F}^{[a_{i_{l}i'_{l}},a_{i_{l}l}]}_{N}(z_{i_{l}},z_{l})}

\newcommand{\gverq} 
{U^{\{r_{q},r_{v_{q}}\}}_{\{(n_{v_{q}})\}}(z_{q})}

\newcommand{\3}{\begin{equation}
\ver\ \ves\ =:\ver\ \ves\ :(z-\xi)^{r\cdot s}\hspace{1cm}for
\:\:\mid\xi\mid < \mid z\mid
\end{equation}}

\newcommand{\lmgcaf}
{{\cal F}^{\{0,0,0,r_{l}\cdot r_{l'}\}}_{N}(z_{l},z_{l'})}

\newcommand{\mgcefm}
{{\cal F}^{[a_{i_{l}i'_{l}},a_{i_{l}l}]-1}_{N}(z_{i_{l}},z_{l})}

\newcommand{\schi}
{{\cal S}^{\{a_{ij},a_{i0},a_{0j},a_{00}\}}_{k_{i},k_{j}}(z_{i},\xi_{j})}

\newcommand{\chia}{
\chi^{\{a_{ij},a_{i0},a_{0j}\}}_{\{(l_{i}+l_{j}),(p_{i}-l_{i}),(p_{j}-l_{j})\}}}

\newcommand{\Wnmderf}{{\cal W}^{n}_{m}}

\newcommand{\Wmnderf}{{\cal W}^{n'}_{m'}}

\newcommand{\wnmderf}{w^{n}_{m}}

\newcommand{\Wmpnderf}{{\cal W}^{n+n'-1}_{m+m'}}

\newcommand{\Vnmderf}{{\cal W}^{1}_{m}}

\newcommand{\Lmwderf}{{\cal L}^{\{a_{i_{l}j_{l'}},a_{i_{l}l'},
a_{l'j_{l'}},a_{ll'}\},\{n_{v_{q}}\}}_{\{n'_{q}\},\{m'_{q}\}}(z_{q})}

\newcommand{\cafl}
{{\cal F}^{\{a_{i_{l}j_{l'}},a_{i_{l}l'},a_{lj_{l'}},a_{ll'}\}}_{ll'}
(z_{i_{l}},z_{j_{l'}})} 

\newcommand{\lschi}
{\vec{\cal S}^{\{a_{ij},a_{i0},a_{0j},a_{00}\}}_{k_{i},k_{j}}(z_{i},\xi_{j})}

\newcommand{\mschi}
{{\cal S}^{\{a_{i_{l}j_{l'}},a_{i_{l}l'},a_{lj_{l'}},a_{ll'}\}}_{\{k_{v_{q}}\}}
(z_{l},z_{l'})}

\newcommand{\Vmnderf}{{\cal W}^{1}_{m'}}

\newcommand{\Vmpnderf}{{\cal W}^{1}_{m+m'}}

\newcommand{\vnmderf}{w^{1}_{m}}

\newcommand{\mwderf}{\partial^{\{n'_{q}\}}_{\{m'_{q}\}}(z_{q})}

\newcommand{\wNMDERF}{\begin{equation}
\wnmderf\ =\sum_{q=1}^{N}z_{q}^{m+n}\frac{\partial^{n}}{\partial z_{q}^{n}}
\end{equation}}

\newcommand{\GEVIR}{\begin{eqnarray}
\vnmderf\ {\cal F}^{\{0,0,0,a_{ll'}\}}_{N}(z_{l},z_{l'})&& =\left [ 
\sum_{l>l'}\frac{a_{ll'}
(z^{m+1}_{l}-z^{m+1}_{l'})}{z_{l}-z_{l'}}\right ]
{\cal F}^{\{0,0,0,a_{ll'}\}}_{N}(z_{l},z_{l'}) \\  
&& =\left [\sum_{l>l'}a_{ll'}
\prod_{k=1}^{m}(z_{l}-\epsilon^{\frac{2k}{m+1}}z_{l'})\right ]
{\cal F}^{\{0,0,0,a_{ll'}\}}_{N}(z_{l},z_{l'}) \nonumber 
\end{eqnarray}}

\newcommand{\CHIS}{\begin{equation}
\chis\ =\sum_{\{\sum_{i}p_{i}=k_{i}, \sum_{j}p_{j}=k_{j}\}}
\sum_{\{l_{i},l_{j}\}}^{\{p_{i},p_{j}\}}\chia\ 
\end{equation}}

\newcommand{\LSCHI}{\begin{eqnarray}
& & \lschi\ =\lim_{\{z_{i}\rightarrow z\}}\lim_{\{\xi_{j}\rightarrow \xi\}}
\schi\ \\ 
& & =\left ( \Delta^{\{k_{i},k_{j}\}}(z_{i},\xi_{j})\gcaf\ \right )
=\frac{\chis\ }
{(z-\xi)^{-r\cdot s+\sum_{i}k_{i}+\sum_{j}k_{j}}} \nonumber
\end{eqnarray}}

\newcommand{\GEVIRZ}{\begin{eqnarray}
w^{1}_{-1} {\cal F}^{\{0,0,0,a_{ll'}\}}_{N}(z_{l},z_{l'})&& =0 \\  
w^{1}_{0} {\cal F}^{\{0,0,0,a_{ll'}\}}_{N}(z_{l},z_{l'})&& =\left [ \sum_{q} 
\frac{1}{2}(\sum_{l\neq q}a_{lq})\right ] 
{\cal F}^{\{0,0,0,a_{ll'}\}}_{N}(z_{l},z_{l'}) \\ 
w^{1}_{1} {\cal F}^{\{0,0,0,a_{ll'}\}}_{N}(z_{l},z_{l'})
&& =\left [ \sum_{q} (\sum_{l\neq q}a_{lq})z_{q}\right ] 
{\cal F}^{\{0,0,0,a_{ll'}\}}_{N}(z_{l},z_{l'}) 
\end{eqnarray}}

\newcommand{\NDELF}{\begin{equation}
\ndelf\ = \prod_{\{q,v_{q}\}}
\lim_{z_{v_{q}}\rightarrow z_{q}}\frac{\partial^{n_{v_{q}}}}{n_{v_{q}}!
\partial z_{v_{q}}^{n_{v_{q}}}}
\end{equation}}

\newcommand{\delf}{\Delta^{\{n_{i},n_{j}\}}(z_{i},\xi_{j})}

\newcommand{\gcaf}
{{\cal F}^{\{a_{ij},a_{i0},a_{0j},a_{00}\}}(z_{i},\xi_{j})} 

\newcommand{\chis}
{\chi^{\{r,s,r_{i},s_{j}\}}_{\{(k_{i}),(k_{j})\}}} 

\newcommand{\mndero}
{\nabla^{\{a_{i_{l}j_{l'}},a_{i_{l}l'},a_{lj_{l'}},a_{ll'}\}}_{\{n_{v_{q}}\}}(z_{v_{q}})}

\newcommand{\ndelf}{\Delta^{\{n_{v_{q}}\}}(z_{v_{q}})}

\newcommand{\nd}
{\nabla^{[a_{i_{l}i'_{l}},a_{i_{l}l}]}_{\{n_{v_{q}}\}}(z_{v_{q}})}

\newcommand{\mgcaf}
{{\cal F}^{\{a_{i_{l}j_{l'}},a_{i_{l}l'},a_{lj_{l'}},a_{ll'}\}}_{N}
(z_{i_{l}},z_{j_{l'}})}

\newcommand{\lmdero}
{\vec{\cal D}^{\{a_{i_{l}j_{l'}},a_{i_{l}l'},a_{lj_{l'}},a_{ll'}\}}_{\{n_{v_{q}}\}}
(z_{v_{q}})}

\newcommand{\lmschik}{\vec{\cal S}^{\{a_{i_{l}j_{l'}},a_{i_{l}l'},
a_{lj_{l'}},a_{ll'}\}}_{\{k_{v_{q}}\}}(z_{l},z_{l'})}

\newcommand{\kdelf}{\Delta^{\{k_{v_{q}}\}}(z_{v_{q}})}

\newcommand{\ND}{\begin{equation}
\nd\ = \ndelf\ \mgcefm\ 
\end{equation}}

\newcommand{\chil}
{\chi^{\{r_{l},r_{l'},r_{i_{l}},r_{j_{l'}}\}}_{\{(k^{l'}_{i_{l}}),
(k^{l}_{j_{l'}})\}ll'}} 

\newcommand{\nwderf}{\partial^{\{n_{q}\}}_{\{m_{q}\}}(z_{q})}

\newcommand{\Lnwderf}{{\cal L}^{\{a_{i_{l}j_{l'}},a_{i_{l}l'},
a_{l'j_{l'}},a_{ll'}\},\{n_{v_{q}}\}}_{\{n_{q}\},\{m_{q}\}}(z_{q})}

\newcommand{\amp}
{{\cal A}^{\{r_{q},r_{v_{q}}\}}(z_{v_{q}},z_{q})}

\newcommand{\gamp}{
G^{\{r_{q},r_{v_{q}}\}}_{N\{(n_{v_{q}})\}}(z_{q})}

\newcommand{\AMP}{\begin{equation}
\amp\ = <:\prod_{q}U^{\{r_{q},r_{v_{q}}\}}(z_{q},z_{v_{q}}):>
=\delta_{\sum_{q}r_{q},0}
\end{equation}}

\newcommand{\lmschikm}{\vec{\cal S}^{\{a_{i_{l}j_{l'}},a_{i_{l}l'},
a_{lj_{l'}},a_{ll'}\}-1}_{\{k_{v_{q}}\}}(z_{l},z_{l'})}

\newcommand{\LNWDERF}{\begin{equation}
\Lnwderf\ =\lmschin\ \nwderf\ \lmschinm\ \end{equation}}

\newcommand{\WGEN}{\begin{eqnarray}
&& [\Lnwderf\ ,\Lmwderf\ ] \\ && =\lmschin\ [\nwderf\ ,\mwderf\ ]\lmschinm\ 
\nonumber \label {eq: WGEN}
\end{eqnarray}}

\newcommand{\EQGT}{\begin{equation}
\Lnwderf\ \gamp\ =0 ~~~~~~for ~~\{n_{q}\neq 0\} \label {eq: EXIT}
 \end{equation}}

\newcommand{\lmschin}{\vec{\cal S}^{\{a_{i_{l}j_{l'}},a_{i_{l}l'},
a_{lj_{l'}},a_{ll'}\}}_{\{n_{v_{q}}\}}(z_{l},z_{l'})}

\newcommand{\lmschinm}{\vec{\cal S}^{\{a_{i_{l}j_{l'}},a_{i_{l}l'},
a_{lj_{l'}},a_{ll'}\}-1}_{\{n_{v_{q}}\}}(z_{l},z_{l'})}

\newcommand{\WINF}{\begin{eqnarray}
[\Wnmderf\ ,\Wmnderf\ ] &&=(n'm-nm')\Wmpnderf\ + \dots \nonumber \\
&&= \sum_{l=1} q^{l-1}C(l)_{n,n'}^{m,m'} W_{m+m'}^{n+n'-l} 
\nonumber \end{eqnarray}
where $q$ is a quantum deformation that is $q=1$ in this case, so:
\begin{equation}
C(l)_{n,n'}^{m,m'} = \left (\begin{array}{cc} n \\ l \end{array}\right )
 \left [\begin{array}{c} m' + n' \end{array}\right ]_{l} - 
\left (\begin{array}{cc} n' \\ l \end{array}\right ) 
\left [\begin{array}{c} m + n \end{array}\right ]_{l} 
\end{equation}}

\begin{document}

\thispagestyle{empty}

\hfill DSF-T-95/49

$ \vspace{1cm} $ 
 
\begin{center}
{\LARGE\bf From Vertex Operators to Calogero-Sutherland Models}
\end{center}

~~~~
\begin{center}
{ V. Marotta $\hspace{.5cm}$  A. Sciarrino } \\
~\\ 
{ Dipartimento di Scienze Fisiche  \\
 Universit\'a di Napoli ``Federico II'' \\ 
and \\ 
INFN, Sezione di Napoli }

\end{center}
~~~~~

\begin{abstract}
\large 

The correlation function of the product of N generalized vertex operators
satisfies an infinite set of Ward identities, related to a $W_{\infty}$ 
algebra, whose extension out of the 
mass shell gives rise to equations which can be
considered as a generalization of the compactified Calogero-Sutherland 
(CS) hamiltonians.
In particular the wave function of the ground state of the compactified CS
model is shown to be given by the value of the product of N vertex
operators between the vacuum and an excited state and the hamiltonian is 
identified with $W^{2}_{0}$ generator. 
The role of the vertex algebra as underlying unifying structure is
pointed out.

\end{abstract}



\vfill
{\small\bf
\begin{tabbing}
Postal address: Mostra d'Oltremare Pad.19-I-80125 Napoli, Italy \\ 
E:mail: \=  Bitnet VMAROTTA (SCIARRINO)@NA.INFN.IT \\ 
        \>  Decnet AXPNA1::VMAROTTA (SCIARRINO)
\end{tabbing}}

\pagebreak

\bigskip

\section{ Introduction }

Recently there has been a renewal of interest in the one dimensional integrable
models, in particular in Calogero-Sutherland (CS) models \cite{14,15}. 
These models describe quantum mechanical systems of N one dimensional 
particles interacting through specific two-body  potentials. 
Although they have been proposed and completely solved in the beginning of the 
seventy, they are at present object of intensive study for their unexpected 
connections with other 
fashionable models as the matrix models where the eigenvalues of the matrix
are identified with the momenta of the particles \cite{22a} and the $c=1$ 
conformal field theory (CFT) models \cite{24} and for their interesting and 
hidden infinite symmetry structure.
The connection of these models with the Lie algebras root systems has been
clarified just a few years after they were proposed and their complete 
integrability was proved \cite{16}. An alternative proof of the
integrability as well as a new expressions of the quantum integrals of the
Sutherland model is given in \cite{17} using representation theory of
Lie algebra $ gl(N) $ and of its affine extension. Only recently the
underlying infinite Kac-Moody symmetry has been understood and their invariance 
for a  $ W_{\infty} $ discovered. In \cite{21} it has been proven that
the Calogero model is invariant for $U(1)$ Kac-Moody, the $U(1)$ current
algebra appearing as the generating function of the quantum integrals of
the model, and for a $ W_{1+\infty} $ algebra. In the case in which the
interacting particles transform as the fundamental representation of a
$ su(N) $ the symmetry is extended by an affine $ su(N) $ and the
$ W_{1+\infty} $ algebra becomes a coloured one. This symmetry has been
further clarified in \cite{23}, where it is also shown that
the Sutherland model is invariant only for the Yangian $ Y(su(N)) $. 
In \cite{22b} it has been proven that the ground state of
the models satisfies the Knizhnik-Zamolodchikov equation.
Moreover using the $S_N$-extended Heisenberg algebra \cite{22c} an
universal Hamiltonian for the CS model has been written as the
anticommutator of one particle operators whose
commutation relations reduce to the standard bosonic 
annihilation-creation operators in the limit of vanishing coupling
constant \cite{22d}. In this approach therefore the potential is
connected with the statistics of the particles.

In this work we show that the very deep structure of these
models, more exactly of their compactified version, i.e. the Sutherland 
model, is the vertex algebra structure,
which constitutes the mathematically rigorous formulation of the algebraic
origin of conformal field theory. We refer for an exhaustive discussion of
the subject to \cite{19} where references to the original papers can
be found. For a pedagogical introduction for physicist see \cite{5}.

The generalized vertex operators provide a well known explicit construction 
of the
vertex algebra and they are connected with the structure of Lorentzian
algebras \cite{7,6,11} which includes the indefinite Kac-Moody algebras
\cite{1} and the Borcherds algebras \cite{4}, the affine algebras being
only a particular and peculiar subset of these algebras.

Even if some of the topics we discuss here can be found scattered in
the literature, we believe useful to present here in a unified matter.
In fact we believe that our approach allows an unified discussion of
the algebraic structure, the connection with conformal field theory and
the statistics of the particles.
The paper is organized as it follows: in Sec. 2  we write the product of
N generalized vertex operators (GVO) in terms of an ordered product of
the GVO times derivative of Jastrow like functions and introduce a set of
differential operators in terms of which we can get all the relevant results of
the vertex algebra. 

In Sec. 3 we define the amplitude or correlation function for the GVOs 
product and show that it can be obtained by the
action of differential operators applied to the amplitude of N standard vertex
operators (VO). 

In Sec. 4 an (infinite) set of Ward identities are then derived 
for the amplitudes, it is shown that the generators spanning a 
$ W_{\infty} $ algebra are a subset of the set of GVO symmetries.  

In Sec. 5 we consider explicitly the particular case of the correlation
functions for VOs and  we show that the wave functions of the ground
state of the CS 
model are obtained as the matrix element of the product of N VO between the 
vacuum and a suitable excited state, the Hamiltonian of the model appearing 
as a combination of differential operators appearing in the analogous of the 
Ward identities for off-shell amplitudes.
Finally in Sec. 6 we briefly discuss further 
developments of the formalism previously developed which in the most general 
case gives rise to more general systems than the CS model.
Let us remark that the vertex operators for the CS models already 
introduced in the literature \cite{25}, can be obtained by expansion 
of the GVOs. 

\section {Differential operators for GVOs}

In this section we introduce a set of differential operators that realize 
the vertex algebra in a complementary way to the usual vertex operators 
construction \cite{19}.

Firstly we briefly recall the GVOs construction, which provides a well known 
realization of vertex algebra \cite{8} and then we show that it is possible to build up
a set of differential operators by means of which to carry out the relevant 
operations on GVOs.

Let us recall the particular choice of basis that we use in our construction 
that simplifies greatly the formal calculations.

The standard (tachyonic) vertex operator (VO) is defined by 
$\ver\ =:e^{ir\cdot Q(z)}:$ where $ Q(z) $ are standard Fubini-Veneziano fields

\1\ 

on d-dimensional Minkowskian torus with periodical boundary conditions given 
by a vector $r$ in a lattice $\Lambda$.

A general GVO is a product of Schur polynomials in the derivatives of Fubini 
fields times a standard vertex operator 
\begin{equation}
\gver\ =:\prod_{i}{\cal P}^{r_{i}}_{n_{i}}(z) \ver\ : = 
:\Delta^{\{n_{i}\}}(z_{i})\prod_{i}U^{r_{i}}(z_{i})U^{r'}(z): 
\end{equation}

where $r'=r-\sum_{i}r_{i}$ and:

\begin{equation}
\Delta^{\{n_{i}\}}(z_{i})=\prod_{i}\lim_{z_{i}\rightarrow z}
\frac{\partial^{n_{i}}}{n_{i}!\partial z_{i}^{n_{i}}} 
\end{equation}

At this point a few words of warning are necessary: in the above equation
the label $z_{i}$ stands for a finite set of variables; to be rigorous 
we should put it in curl brackets. However in order not to overweight the
notation we use only the round brackets. We shall use this convention
hereafter. Only for the GVO or VO the functional dependance is only from
one variable. Moreover in the above definition, we should also write the
explicit dependence on the $z$ point where the limit is done at the end. For
the same reason we shall omit here and in similar equations the dependence
on the point(s) where the limit is (are) performed.

Formal Laurent expansion of GVO is denoted as:
\begin{equation}
\gver\ =\sum_{m} A^{\{r,r_{i}\}}_{\{(n_{i})\}m}z^{-m-h}
\end{equation}
where $h=\frac{r^{2}}{2}+\sum_{i}n_{i}$ is the conformal weight.

The product of two VOs is simply:
\3\ 
by means of locality properties the product is analytically extended to whole 
${\cal C}^{2}$ space except the poles $z=\xi$ and $z,\xi=0,\infty $.

We can easily generalize this relation to $N$ VOs product:
\begin{equation}
\prod_{q=1}^{N}U^{r_{q}}(z_{q})=:\prod_{q=1}^{N}U^{r_{q}}(z_{q}):\lmgcaf\ 
\end{equation}

where we have introduced a Jastrow like function:
\begin{equation}
\lmgcaf\ = \prod_{l>l'} (z_{l}-z_{l'})^{r_{l}\cdot r_{l'}}
\end{equation}

Using this result and the differential operators:

\begin{equation}
\delf\ = \prod_{i,j}\lim_{z_{i}\rightarrow z}\lim_{\xi_{j}\rightarrow \xi}
\frac{\partial^{n_{i}}}{n_{i}!\partial z_{i}^{n_{i}}}
\frac{\partial^{n_{j}}}{n_{j}!\partial \xi_{j}^{n_{j}}} \label {eq: DELTA}
\end{equation}

we can compute any two GVOs product and we get:
\begin{equation}
\gver\ \gves\ = \delf\ :U^{\{r,r_{i}\}}(z,z_{i})U^{\{s,s_{j}\}}(\xi,\xi_{j}): 
\gcaf\ \label {eq: GVOPROD}
\end{equation}
where 
\begin{eqnarray}
&& U^{\{r,r_{i}\}}(z,z_{i})=:\prod_{i}U^{r_{i}}(z_{i})U^{r'}(z): ~~~~ 
r'=r-\sum_{i}r_{i} 
\\ 
&& U^{\{s,s_{j}\}}(\xi,\xi_{j})=:\prod_{j}U^{s_{j}}(\xi_{j})U^{s'}(\xi): 
~~~~ s'=s-\sum_{j}s_{j}
\end{eqnarray}
and
\begin{equation}
\gcaf\ 
=\prod_{ij}(z_{i}-\xi_{j})^{a_{ij}}(z_{i}-\xi)^{a_{i0}}(z-\xi_{j})^{a_{0j}}
(z-\xi )^{a_{00}} 
\end{equation}
with $a_{ij}=r_{i}\cdot s_{j}$, $a_{0j}=r'\cdot s_{j}$, $a_{i0}=r_{i}\cdot s' $
 e $ a_{00}=r'\cdot s' $.

After some algebraic calculation we obtain:
\begin{equation}
\gver\ \gves\ = \sum^{n_{i},n_{j}}_{k_{i},k_{j}}\frac{\chis\ }
{(z-\xi)^{-r\cdot s+\sum_{i}k_{i}+\sum_{j}k_{j}} }
:U^{\{r,r_{i}\}}_{\{(n_{i}-k_{i})\}}(z)U^{\{s,s_{j}\}}_{\{(n_{j}-k_{j})\}}(\xi):
\end{equation}

with
\begin{equation}
\chis\ =\sum_{\{\sum_{i}p_{i}=k_{i}, \sum_{j}p_{j}=k_{j}\}}
\sum_{\{l_{i},l_{j}\}}^{\{p_{i},p_{j}\}}
(-1)^{\sum_{j}l_{j}}\prod_{i,j}
\left (\begin{array}{cc} a_{i0} \\p_{i}-l_{i}\end{array}\right )
\left (\begin{array}{cc} a_{ij} \\l_{i}+l_{j}\end{array}\right )
\left (\begin{array}{cc} a_{0j}\\p_{j}-l_{j}\end{array}\right )  
\end{equation}

Remark that this relation is obtained in a non-local way without OPE expansion 
so it holds also 
in the large distances limit and can be used to give non perturbative results.

The set of $a_{ll'}$ can be expressed in terms of upper triangular matrices in 
$gl(N)$ so relating our approach to the approach of ref.\cite{17}.

To extend eq.(\ref {eq: GVOPROD}) we remark that only Jastrow like functions appear
in the GVOs product so we can define:
\begin{equation}
\prod_{q}^{N}\gverq\ =\ndelf\ :\prod_{q}^{N}U^{\{r_{q},r_{v_{q}}\}}
(z_{q},z_{v_{q}}):\mgcaf\ \label {eq: MGVOP} 
\end{equation}
where:
\NDELF\ 
is the generalization of operators in eq.(\ref {eq: DELTA}) and
\begin{equation}
\mgcaf\ = \prod_{l>l'}\prod_{i_{l}j_{l'}}(z_{i_{l}}-z_{j_{l'}})^{a_{i_{l}j_{l'}}}
(z_{i_{l}}-z_{l'})^{a_{i_{l}l'}}(z_{l}-z_{j_{l'}})^{a_{lj_{l'}}}
(z_{l}-z_{l'})^{a_{ll'}} 
\end{equation}

\begin{equation} 
:\prod_{q}^{N}U^{\{r_{q},r_{v_{q}}\}}(z_{v_{q}},z_{q}): 
= :\prod_{q}^{N}\prod_{\{v_{q}\}}U^{r_{v_{q}}}(z_{v_{q}})U^{r'_{q}}(z_{q}):
\end{equation}

From these definitions we can introduce differential operators that acting only 
on normal ordered products of tachyonic vertex give expression for general GVOs:
\begin{equation}
\lmdero\ = \ndelf\ \mgcaf\ 
\end{equation}

If we expand these operators in terms of $\ndelf\ $ we obtain:
\begin{equation}
\lmdero\ =\sum^{\{n_{v_{q}}\}}_{\{k_{v_{q}}\}}\lmschik\ 
\Delta^{\{n_{v_{q}}-k_{v_{q}}\}}(z_{v_{q}})
\end{equation}
where
\begin{equation}
\lmschik\ = \left (\kdelf\ \mgcaf\ \right )
\end{equation}

These functions can be explicitly computed, see the Appendix, and have the 
following expression:
\begin{equation} 
\lmschik\ = 
\sum_{\{\sum_{l>q} k_{v_{q}}^{l}+\sum_{l'<q} k_{v_{q}}^{l'}=k_{v_{q}}\}}
\prod_{l>l'}
\frac{\chil\ }
{(z_{l}-z_{l'})^{-r_{l}\cdot r_{l'}+\sum_{i_{l}}k_{i_{l}}^{l'}
+\sum_{j_{l'}}k_{j_{l'}}^{l}}} 
\end{equation}

The explicit expansion of general $N$ GVOs product becomes:
\begin{eqnarray}
&&\prod_{q}^{N}U^{\{r_{q},r_{v_{q}}\}}_{\{(n_{v_{q}})\}}(z_{v_{q}},z_{q})= 
\lmdero\ :\prod_{q}^{N}U^{\{r_{q},r_{v_{q}}\}}(z_{v_{q}},z_{q}): 
\\ &&=
\sum_{\{\sum_{l>q} k_{v_{q}}^{l}+\sum_{l'<q} k_{v_{q}}^{l'}=k_{v_{q}}\}}
\prod_{l>l'}
\frac{\chil\ }
{(z_{l}-z_{l'})^{-r_{l}\cdot r_{l'}+\sum_{i_{l}}k_{i_{l}}^{l'}
+\sum_{j_{l'}}k_{j_{l'}}^{l}}} 
:\prod_{q}^{N}U^{\{r_{q},r_{v_{q}}\}}_{\{(n_{v_{q}}-k_{v_{q}})\}}
(z_{v_{q}},z_{q}): \nonumber 
\end{eqnarray}

Locality properties for GVOs can be evaluated by means of permutations group 
(notice that $\ndelf\ $ and ordered products are $S_{N}$-symmetric so we do 
not indicate explicitly the symmetrization): 
\begin{eqnarray}
&& \Pi_{qq'}\prod_{q}^{N}\gverq\ \Pi_{qq'} = \\ 
&&=\ndelf\ :\prod_{q}^{N}U^{\{r_{q},r_{v_{q}}\}}(z_{q},z_{v_{q}}):K_{qq'}
\mgcaf\ \nonumber    
\end{eqnarray}
where $\Pi_{qq'}$ is an operator that exchanges the order of two GVOs while 
$K_{qq'}$ 
exchanges the sets of indices $\{q, v_{q}\}$ and $\{q', u_{q'}\}$ in the 
functions: 
\begin{equation}
K_{qq'}\mgcaf\ =\epsilon ^{\sum_{l=0}^{q'-q-1}r_{q+l}\cdot r_{q'}}\mgcaf\ 
\end{equation}
where $\epsilon = e^{i \pi} $ so we generalize this construction also to 
non-local cases (in particular rational values of $ r_{q}\cdot r_{q'} $ 
give RCFT) where vertex algebras can be extended following \cite{30}.

\bigskip

\section{GVOs correlators}

By means of this formalism we calculate explicitly also correlation functions 
for GVOs using the property of normal ordered VOs:
\AMP\
so by using of eq.(\ref {eq: MGVOP}) we have simply:
\begin{eqnarray}
\gamp\ && = <\prod_{q}U^{\{r_{q},r_{v_{q}}\}}_{\{(n_{v_{q}})\}}(z_{q})> 
=\lmdero\ \amp\ \nonumber \\ 
&& = \lmschin\ \delta_{\sum_{q}r_{q},0} 
\end{eqnarray}

We can also verify the duality invariance by using of action of symmetric 
group:
\begin{equation}
\prod_{l\neq i}K_{il} \gamp\ = \prod_{l\neq i}\eta_{il}(r_{i}\cdot r_{l})
\gamp\  = \epsilon^{-r_{i}^{2}}\gamp\ ~~\forall ~ i 
\end{equation}
where $\eta_{il}(r_{i}\cdot r_{l})=\epsilon^{r_{l}\cdot r_{i}}$ and  
neutrality implies $\sum_{l\neq i}r_{l}\cdot r_{i}=-r_{i}^{2}$.

A $\gamp\ $ amplitude can be obtained also by means of the action of the 
$\ndelf\ $ operators on a tachyonic amplitude:
\begin{equation}
\gamp\ = \nd\ G^{\{r_{q},r_{v_{q}}\}}_{N+\sum n_{v_{q}}}(z_{v_{q}},z_{q})
\end{equation}
where
\ND\ 
\begin{equation}
\mgcef\ = \prod_{l}\prod_{i_{l}i'_{l}}(z_{i_{l}}-z_{i'_{l}})^{a_{i_{l}i'_{l}}}
(z_{i_{l}}-z_{l})^{a_{i_{l}l}}
\end{equation}
in fact applying this operator we obtain:
\begin{equation} 
\nd\ G^{\{r_{q},r_{v_{q}}\}}_{N+\sum n_{v_{q}}}(z_{v_{q}},z_{q})= 
\lmschin\ \delta_{\sum_{q}r_{q},0} 
\end{equation}

So all ``massive" amplitudes properties can be deduced simply by tachyonic ones.

This set of operators gives relations between $N$  and 
$N+\sum n_{v_{q}}$ vertex amplitudes 
so they appear as non-linear realization of a larger algebra 
as pointed out by Witten \cite{27} in the case of 2D string theory.

By using of a diverse factorization of functions:
\begin{equation}
\mgcaf\ =
{\cal F}^{\{a_{i_{l}j_{l'}},a_{i_{l}l'},a_{lj_{l'}},a_{ll'}-r_{l}\cdot r_{l'}\}}_{N}
(z_{i_{l}},z_{j_{l'}}){\cal F}^{\{0,0,0,r_{l}\cdot r_{l'}\}}_{N}(z_{l},z_{l'}) 
\label {eq: FATMAS}
\end{equation}
we obtain also a relation between higher spin ``excitations" and 
tachyonic amplitudes with the same momentum:
\begin{eqnarray}
&& \gamp\ =\nd\ G^{\{r_{q},r_{v_{q}}\}}_{N}(z_{v_{q}},z_{q}) \\ && = 
\mndero {\cal F}^{\{a_{i_{l}j_{l'}},a_{i_{l}l'},a_{lj_{l'}},a_{ll'}-r_{l}\cdot r_{l'}\}}_{N}
(z_{i_{l}},z_{j_{l'}}){\cal F}^{\{0,0,0,r_{l}\cdot r_{l'}\}}_{N}
(z_{l},z_{l'}) \delta_{\sum_{r_{q}},0} \nonumber \\ 
&& =\vec{\cal S}^{\{a_{i_{l}j_{l'}},a_{i_{l}l'},a_{lj_{l'}},a_{ll'}-
r_{l}\cdot r_{l'}\}}_{\{n_{v_{q}}\}}(z_{v_{q}})G^{\{r_{q}\}}_{N}(z_{q}) 
\nonumber 
\end{eqnarray}

An important point in this approach is that we can built all amplitudes 
by means of the action of $\ndelf\ $ set of differential operators 
on the tachyonic correlation functions. 

Moreover, all informations on vertex algebra are contained in the contact 
${\cal S}$ functions so we can remove completely the GVOs from our construction and make 
use only of differential operators and $\mgcaf\ $ functions to study any
amplitudes.

\bigskip

\section{ Ward identity for correlation functions }

 At this point we can study the whole set of amplitudes in order to recover
their symmetries. To do this we note that if the sum of roots vanishes 
({\bf mass shell condition}) ordered amplitudes $\amp\ $ are 
constant or in a more general case are symmetric functions,  
so they satisfy an infinity of differential equations:
\begin{equation}
\nwderf\ \amp\ = \left (\prod_{q}z_{q}^{m_{q}+n_{q}}\frac{\partial^{n_{q}}}{
n_{q}!\partial z_{q}^{n_{q}}}\right ) \amp\ =0
\end{equation}

In the language of CFT these relations correspond to the insertion of 
quasi-primary fields \cite{11}, in fact:
\begin{equation}
\nwderf\ \amp\ = <:\prod_{q}z_{q}^{m_{q}+n_{q}}\frac{\partial^{n_{q}}}{
n_{q}!\partial z_{q}^{n_{q}}}\gverq\ :>
\end{equation}
is the ordered amplitude for descendent fields relatives to SL(2,R) 
subalgebra.

To these ordered amplitudes symmetries we associate Ward identities for 
generic GVOs amplitudes, which, defining the operators:
\LNWDERF\

can be written in the following form:
\EQGT\

The above introduced operators satisfy the same commutation relations of 
free differential algebra $\nwderf\ $:
\WGEN\

and the Jacobi identity.

Remark that the function $\lmschikm\ $ is a sum of Jastrow functions and is 
well defined everywhere except in the points $z_{l}=z_{l'}$ but 
in these points the singularity of GVOs are regularized with normal ordering.

Using the factorization given in eq.(\ref {eq: FATMAS}) it is possible also 
to write a set of effective equations for the ``excitations", which, 
introducing the following operators:
\begin{eqnarray}
&& {\cal L}^{\{a_{i_{l}j_{l'}},a_{i_{l}l'},
a_{l'j_{l'}},a_{ll'}-r_{l}\cdot r_{l'}\},\{n_{v_{q}}\}}_{
\{n_{q}\},\{m_{q}\}}(z_{q}) \\ && =
{\cal F}^{\{0,0,0,r_{l}\cdot r_{l'}\}-1}_{N}(z_{l},z_{l'})
{\cal L}^{\{a_{i_{l}j_{l'}},a_{i_{l}l'},
a_{l'j_{l'}},a_{ll'}\},\{n_{v_{q}}\}}_{\{n_{q}\},\{m_{q}\}}(z_{q})
{\cal F}^{\{0,0,0,r_{l}\cdot r_{l'}\}}_{N}(z_{l},z_{l'}) \nonumber 
\end{eqnarray}
can be written as:
\begin{equation}
{\cal L}^{\{a_{i_{l}j_{l'}},a_{i_{l}l'},
a_{l'j_{l'}},a_{ll'}-r_{l}\cdot r_{l'}\},\{n_{v_{q}}\}}_{
\{n_{q}\},\{m_{q}\}}(z_{q})\vec{\cal S}^{\{a_{i_{l}j_{l'}},a_{i_{l}l'},
a_{lj_{l'}},a_{ll'}-r_{l}\cdot r_{l'}\}}_{\{n_{v_{q}}\}}(z_{v_{q}})=0
~~~for ~~~\{n_{q}\neq 0\} 
\end{equation}
The above equation are the analogous of eq.(\ref{eq: EXIT}) for $S$ functions
and satisfies the same algebraic relations. 

The set of differential operators introduced in eq.(38) can be considered as 
a generalization of Dunkl 
operators whose role is fundamental in the theory of integrable systems. 

In fact from the property:
\begin{eqnarray}
&& \lmschin\ \nwderf\ \lmschinm\   \\ && =
\prod_{q}\frac{z_{q}^{m_{q}+n_{q}}}{n_{q}!}
\left ( \lmschin\ \frac{\partial}{\partial z_{q}} 
\lmschinm\ \right )^{n_{q}} \nonumber 
\end{eqnarray}
, using the identity:
\begin{eqnarray}
&& \lmschin\ \frac{\partial}{\partial z_{q}} \lmschinm\  \\ && =-
\lmschinm\ \frac{\partial}{\partial z_{q}} \lmschin\ \nonumber 
\end{eqnarray}
we can write a general differential operator in a more simple way:
\begin{equation}
\Lnwderf\ =(-1)^{\sum_{q}n_{q}}\prod_{q}\frac{z_{q}^{m_{q}+n_{q}}}{n_{q}!}
\left (\hat{\cal L}^{\{a_{i_{l}j_{l'}},a_{i_{l}l'},
a_{l'j_{l'}},a_{ll'}\},\{n_{v_{q}}\}}_{\{1_{q}\}}(z_{q})\right )^{n_{q}}
\end{equation}
The generalized Dunkl operator is:
\begin{equation}
\hat{\cal L}^{\{a_{i_{l}j_{l'}},a_{i_{l}l'},
a_{l'j_{l'}},a_{ll'}\},\{n_{v_{q}}\}}_{\{1_{q}\}}(z_{q})
=-\lmschinm\ \frac{\partial}{\partial z_{q}} \lmschin\ 
\end{equation}

which in fact reduces to the usual Dunkl operator in the case of VOs:
\begin{eqnarray}
\hat{\cal L}^{\{0,0,0,a_{ll'}\}-1}_{\{1_{q}\}}(z_{q})
&& =-{\cal F}^{\{0,0,0,r_{l}\cdot r_{l'}\}}_{N}(z_{l},z_{l'}) 
\frac{\partial}{\partial z_{q}} {\cal F}^{\{0,0,0,r_{l}\cdot r_{l'}\}}_{N}
(z_{l},z_{l'}) \\ 
&& =-\frac{\partial}{\partial z_{q}}+\sum_{l\neq q}\frac{a_{lq}}{z_{l}-z_{q}} 
 \nonumber \label {eq: 46}
\end{eqnarray}

Moreover, in this case we have:
\begin{equation}
\hat{\cal L}^{\{0,0,0,a_{ll'}\}-1}_{\{1_{q}\}}(z_{q}) 
G^{\{r_{i}\}}_{N}(z_{i})   = 
<(\frac{\partial}{\partial z_{q}} - r_{q}\cdot Q^{(1)}(z_{q}))\prod_{i}^{N}
U^{r_{i}}(z_{i})>
\end{equation}
where the singular product is understood as limit $z\rightarrow z_{q}$.

So in this case we can identify the action of the Dunkl operator on the VOs 
amplitudes with the action of the operator:
\begin{equation}
\frac{\partial}{\partial z_{q}} - r_{q}\cdot Q^{(1)}(z_{q})
\end{equation}
on the product of VOs that is just a Miura transformation generating an 
explicit realization of $W_{\infty}$ in terms of fields.

In the general case of $N$ GVOs, however, a Miura transformation cannot exist.

\bigskip

\subsection{ $W_{\infty}$ symmetries of amplitudes }

In this section we describe some interesting subalgebras of the GVOs 
symmetries by using the correspondence between ordered amplitudes $\amp\ $ 
and $\gamp\ $ which allows us to single out the relevant symmetries.

It is obvious that the subalgebra of $\nwderf\ $ algebra spanned by
the generators:
\wNMDERF\
satisfies a $W_{\infty}$ algebra.

The corresponding differential operators $\Wnmderf\ $ obtained replacing
$\nwderf\ $ with the operators of eq.(\ref {eq: WGEN}) are a realization 
of the same algebra.

An important consequence of this is that all GVOs amplitudes are $W_{\infty}$ 
invariant, as indeed the $\Wnmderf\ $ generators satisfy quantum commutation 
relations without central extensions,, this is an obvious consequence of zero genus of Riemann surface on which we take 
correlators, while anomalies arise on genus one surface:
\WINF\
and $[a]_{b}$ is the Pochhammer symbol.

It should be possible to realize these generators by means of a particular 
combination of GVOs.

In particular we compute the Virasoro (Witt) subalgebra generators:
\begin{equation}
\left [\Vnmderf\ , \Vmnderf\ \right ]= (m-m')\Vmpnderf\
\end{equation}

In the case of Virasoro algebra these vertices can be simply identified.
In fact in this case, it is well know that the tensor field $T(\xi)$ is the 
generator of transformations that act on VOs in the following way:
\begin{equation}
\oint_{C_{\xi,z}} \frac{d\xi }{2 \pi i} \xi^{m+2}T(\xi)\ver\ = 
\left ( z^{m+1}\frac{\partial}{\partial z} + \frac{r^{2}}{2}(m+1)z^{m} \right )
\ver\ 
\end{equation}
where the circuit $C_{\xi,z}$ includes the pole $\xi= z$.

If we specialize these operators to the projective subalgebra we obtain:
\GEVIRZ\
where $\sum_{l\neq q}a_{lq}=-r^{2}_{q}=-2h_{q}$.

So the invariance is satisfied with the trivial comultiplication and we can 
use these relations to fix three variables in $\lmschin $ functions.

In this case it is simple to realize these transformations by means of $T(\xi )$:
\begin{equation}
\Vnmderf\ G^{\{r_{q}\}}_{N}(z_{q}) = \sum_{q}\oint_{C_{\xi,z_{q}}} \frac{d\xi }{2 \pi i}\xi^{m+2}
<U^{r_{1}}(z_{1})\cdots T(\xi)U^{r_{q}}(z_{q})\cdots U^{r_{N}}(z_{N})>=0
\end{equation}

In the general case additional terms are needed:
\GEVIR\
In fact as the vacuum state $\mid 0>$ is invariant only for 
projective transformations, in this case it is necessary to consider 
also the contribution of the action of $T(\xi)$ on the vacuum:
\begin{eqnarray}
\Vnmderf\ G^{\{r_{q}\}}_{N}(z_{q}) 
&& = \sum_{q}\oint_{C_{\xi,z_{q}}} \frac{d\xi }{2 \pi i}\xi^{m+2}
<U^{r_{1}}(z_{1})\cdots T(\xi)U^{r_{q}}(z_{q})\cdots U^{r_{N}}(z_{N})> 
\nonumber \\ 
&& +\oint_{C_{\xi,0}} \frac{d\xi }{2 \pi i}\xi^{m+2}
<U^{r_{1}}(z_{1})\cdots U^{r_{N}}(z_{N})T(\xi)> \nonumber \\
&& +\oint_{C_{\xi,\infty}} \frac{d\xi }{2 \pi i}\xi^{m+2}
<T(\xi)U^{r_{1}}(z_{1})\cdots U^{r_{N}}(z_{N})>=0
\end{eqnarray}

The interpretation of  this result is very simple: the $W_{\infty}$ 
symmetry of GVOs in quantum case is realized taking in to account the anomalous 
transformations of the vacuum, so the symmetry is restored also in non-critical 
dimensions.

\bigskip

\subsection{ Additional  equations}

In the previous section it is shown that a $W_{\infty}$ algebras exists for any 
GVOs correlator and the explicit realization can be given in terms of 
generalized Dunkl operators.

This is only a subalgebra of the full symmetry algebra that can be realized, in 
this section we want to understand the role of remaining differential operators.

As recently is pointed out \cite{33}, canonical quantization of two dimensional 
identical particles give unusual interesting results.

We consider GVOs correlator as multiparticle form factors and the generalized 
Dunkl operators as one-particles operators, in this framework the 
$W_{\infty}$ generators become the completely symmetric single-particles 
operators that give the observables of $N$ particles system.

We can also construct operators not $S_{N}$ invariant that must be  
related to the differential equations not in $W_{\infty}$ algebra.

Two cases can arise: in the first the particles are distinguishable then these 
operators give observable results; in the second case the particles are 
identical then  
they cannot be observable to avoid the possibility to identify the particle.

In this framework the existence of this additional symmetry have a very simply 
physical interpretation in terms of breaking the Hilbert space of the  
particles in $N!$ sectors that cannot be mixed by hamiltonian evolution of the 
system so all not-symmetric operators must be unobservable.

This aspect is very interesting because it implies that must exist many 
null states for $W_{\infty}$ representations corresponding to these 
unobservable operators; this is just the case of quasi-finite representations 
that are well studied in \cite{31}. 

Notice that degenerate representations of W algebras arise also in the 
classification of hierarchies in quantum Hall effect \cite{34} without any 
explicit request of generalized exclusion principle.

In the following section we explore in more detail this aspect in the case of 
CS model.
 
These further equations relate amplitudes in which there are quasi-secondary fields 
to pure quasi-primary ones.
Therefore, by means of these equations we reduce the vertex operators space 
to quasi-primary states only.
In this way we remove not only the dependence on quasi-secondary null states 
but also on all quasi-secondary as a signal of enhanced symmetry. 
This invariance has a simple interpretation in terms of universal enveloping 
algebra of projective symmetry and does not depend on full Virasoro constraints. 

As pointed out in \cite{11} this is very important for the consistence of GVOs 
construction of Lorentzian algebras. 

Besides the considered linear symmetry algebra of GVOs which commutes with the 
$N$ operator (number of GVOs), we remark that it does exist another set of 
symmetry obtained applying the $\nabla$ operator of Sec.(3) which relate 
product of GVOs with different value of $N$, see the end of Sec.(5).

\bigskip

\section{ Relationship with Sutherland model}

In the first part of this paper we have constructed the differential 
operators that generate the symmetries for all GVOs amplitudes, in what follows 
we describe the connection with the CS one dimensional integrable systems.
The compactification of external space implies that periodic boundary 
conditions for the momentum of particles must be imposed, so the model that we 
describe is of 
Sutherland form that describes a system of non-relativistic particles on a 
circle interacting with an inverse square potential.

This model is completely integrable and gives eigenfunctions expressed in 
terms of Jack polynomials $J^{(r^{2})}_{\{t\}}(z_{q})$ that are indexed by the 
Young diagrams that can be interpreted as the distribution of the momentum 
of pseudo-particles (holes).

The Young diagram is parameterized by the $N$ numbers $\{t\}=(t_{1},\dots 
,t_{N})$, $t_{1}\geq t_{2}\cdots \geq t_{N}\geq 0 $ with total number of boxes 
denoted by $\mid t\mid=\sum_{q=1}^{N}t_{q}$.

To make a reduction of GVOs space we have to impose that the number of 
particles is a 
constant so we consider only the set of $\gamp\ $ with a fixed value of $N$. 
The correspondence with the CS model is obtained by the identification of the
states and of the hamiltonian, which can be done using the projective 
invariance to fix two states corresponding to in-vacuum and out-excited state:
\begin{equation}
\psi^{\{s,s_{j},r_{q},r_{v_{q}},r,r_{i}\}}_{\{(n_{j})(n_{v_{q}})(n_{i})\}N}
(z_{q})= 
<s,s_{j}(n_{j})\mid \prod_{q=1}^{N}\gverq\ \mid r,r_{i}(n_{i})>
\end{equation}

Specializing these wave functions to Sutherland model we consider the ground 
state ($r_{q}=r ,~ r_{v_{q}}=0 ~~\forall ~q$):
\begin{equation}
\psi^{\{r,\lambda\}}_{N}(z_{q})= <Nr+\lambda\mid \prod_{q=1}^{N}U^{r}(z_{q})\mid \lambda> 
= \prod_{l>l'}(z_{l}-z_{l'})^{r^{2}}\prod_{q=1}^{N}z_{q}^{r\cdot \lambda} 
\end{equation}

The wave function vanishes for $ z_{l} -  z_{l'} \rightarrow 0$ if $r^{2} > 0$,
which indeed is the condition ensuring the absence of poles in the product of
two VOs and assure the normalizabilty of the wave functions, it is very 
interesting to notice that we could consider also the case 
$r^{2}<0$ where bound-states should exist.

When expand VOs in terms of symmetric Jack polynomials we obtain the wave 
functions of excited states \cite{5}:
\begin{equation}
:\prod_{q=1}^{N}U^{r}(z_{q}):\mid\lambda>=\prod_{q=1}^{N}z_{q}^{\lambda^{2}}
\sum_{\{t\}} (-1)^{\mid t\mid} J^{(r^{2})}_{\{t\}}(z_{q})
\left (j^{r^{2}}_{t}\right )^{-1/2}\mid \{t\}, Nr+\lambda>
\end{equation}
where $\left (j^{r^{2}}_{t}\right )^{-1/2}$ is a normalization factor.

All informations on symmetry of states are encoded in the ground state wave 
functions, this allows to understand the importance of GVOs in 1-dimensional 
integrable models.
GVOs can be factorized in a Jastrow like function and a symmetric term that 
 depends only on ordered products.
Statistical properties are related only to $\mgcaf\ $ functions.    

By orthogonality properties of Jack polynomials we obtain:
\begin{equation}
<\{t\},Nr+\lambda\mid 
\prod_{q=1}^{N}U^{r}(z_{q})\mid\lambda>=\prod_{l>l'}(z_{l}-z_{l'})^{r^{2}}
\prod_{q=1}^{N}z_{q}^{r\cdot\lambda+\lambda^{2}}
(-1)^{\mid t\mid} J^{(r^{2})}_{\{t\}}(z_{q})
\left ( j^{r^{2}}_{t}\right )^{-1/2}
\end{equation}

The completeness of Jack polynomials in the space of $c=1$ CFT at 
$R=\sqrt{r^{2}}$ implies that exist an out-state for each state of Sutherland 
model, so $W_{\infty}$ symmetry project the functions defined in $R^{N}$ 
into single-particle space, in this way full set of wave functions can be 
obtained acting only on out-state (collective state).

In this case the first two generators of $W_{\infty}$ algebra are:
\begin{equation}
W_0^1 =\prod_{l>l'}(z_{l}-z_{l'})^{r^{2}} \sum_{q}z_{q}\frac{\partial}{\partial z_{q}}
\prod_{l>l'}(z_{l}-z_{l'})^{-r^{2}}= 
\sum_{q}z_{q}\frac{\partial}{\partial z_{q}} - \frac{r^{2}}{2}N(N-1)   
\end{equation}
and
\begin{eqnarray}
W_0^2 &&= \prod_{l>l'}(z_{l}-z_{l'})^{r^{2}}
\sum_{q}z_{q}\frac{\partial^2}{\partial z_{q}^2} 
\prod_{l>l'}(z_{l}-z_{l'})^{-r^{2}} \\ 
&&=\sum_{q} z_{q}^{2}\frac{\partial^{2}}{z_{q}^{2}} - 2 
r^{2}\sum_{l>l'} 
\frac{z_{l}^{2}\frac{\partial}{\partial z_{l}}
- z_{l'}^{2}\frac{\partial}{\partial z_{l'}}}{z_{l}-z_{l'}} +
2 r^{2}(r^{2} + 1)\sum_{l>l'}\frac{z_{l}z_{l'}}{(z_{l}-z_{l'})^{2}} 
\nonumber \\  &&+ r^{4}
\sum_{q,l,l'\neq q}\frac{z_{q}^{2}}{(z_{l}-z_{q})(z_{q}-z_{l'})} + 
\frac{r^{2}(r^{2} + 1)}{2}N(N-1) 
\end{eqnarray}

The first generator corresponds, up to a constant, to total momentum $P$: 
\begin{equation}
P =\sum_{q}z_{q}\frac{\partial}{\partial z_{q}} 
\end{equation}

while the first two terms of the second generator are the differential operators 
whose eigenvectors are, according to Stanley's theorem \cite{38}, the Jack 
polynomials, while the fourth term is a constant.
This suggests that there is a close relationship between this generator and 
the CS hamiltonian which in our notation is written as:
\begin{equation}
H =\sum_{q}\frac{1}{2}\left ( z_{q}\frac{\partial}{\partial z_{q}}\right )^{2}-
r^{2}(r^{2}-1)\sum_{l>l'}\frac{z_{l}z_{l'}}{(z_{l}-z_{l'})^{2}}
\end{equation}

Moreover, by construction on the ground state we have:
\begin{eqnarray}
{\cal W}^{1}_{0}\psi^{\{r,\lambda \}}_{N}(z_{q})&&=
\left [ P- \frac{r^{2}}{2}N(N-1)\right ]\psi^{\{r,\lambda \}}_{N}(z_{q})=
p\psi^{\{r,\lambda \}}_{N}(z_{q}) \\ 
{\cal W}^{2}_{0}\psi^{\{r,\lambda \}}_{N}(z_{q})&&=\left [2H+
\frac{r^{4}}{6}N(N^{2}-1)\right ]
\psi^{\{r,\lambda \}}_{N}(z_{q})=2E \psi^{\{r,\lambda \}}_{N}(z_{q})
\end{eqnarray}
where $p=Nr\cdot\lambda $ and $E=\frac{1}{2}N(r\cdot\lambda)^{2}$ are the 
eigenvalues of relative operators.

In order to be able to really identify the CS hamiltonian we have to modify 
our approach introducing the symmetric hermitian exchange operators 
$K_{ll'}$, so we can discuss the more general case in which the exchange operator
does appear in the CS hamiltonian. 

In the following we give more general results that it is possible to have by 
the VOs construction if $r_l \neq r_{l'}$.

The CS wave function $\Psi(z_q)$ can be written in a factorized form as the 
product of the ground
state wave-function $\psi^{\{r_{q},\lambda \}}_{N}(z_{q})$ times the wave 
function of the collective CS
hamiltonian i.e. for $a_{ll'}=r^{2}$ a Jack polynomial $\phi (z_{q})$. 
If the particles are identical
(indistinguishable) the wave function has to be invariant under the action
of the operator $K_{ll'}$, so we must impose, for any couple $ll'$
\begin{equation}
(1 - \eta_{ll'}K_{ll'}) \Psi(z_q) = \psi^{\{r_{q},\lambda \}}_{N}(z_{q})
 (1 - K_{ll'}) \phi(z_q) = 0
\end{equation}
where $\eta_{ll'}$ is the phase, computed in Sec.(2), produced by the action
of $K_{ll'}$ on $\psi^{\{r_{q},\lambda \}}_{N}(z_{q})$.

The phase $\eta_{ll'}$ takes in account of the statistics that depends on 
length of the $r_{q}$ roots.

Let use remark that the invariance of the wave function should require 
identical particles, i.e. all $r_{q}$ equal or the introduction of mutual 
exclusion statistics \cite{40}, whose connection with CS model has been 
discussed in \cite{41} and \cite{42}. 

The VOs approach naturally leads to models with this type of statistics.

We make use of results in Sec.(4.2) that allow us to replace the differential 
operator of eq.(\ref {eq: 46}) by the Dunkl operator
\begin{equation}
 d_q =
\frac{\partial}{\partial z_{q}} +
\sum_{l\neq q}\frac{a_{ql}}{z_{q}-z_{l}}(1 - K_{ql})
\end{equation}

When $a_{ql}=r^{2}$ the $d_{q}$ satisfy the relations of an affine Hecke 
algebra, but now we have:
\begin{eqnarray}
\left [ d_{q},d_{q'}\right ] &&=\sum_{l\neq q,q'}\left ( 
\frac{a_{qq'}a_{q'l}}{(z_{q}-z_{q'})(z_{q'}-z_{l})}-
\frac{a_{ql}a_{lq'}}{(z_{q}-z_{l})(z_{q'}-z_{l})} \right. \nonumber \\ 
&&\left. - \frac{a_{qq'}a_{ql}}{(z_{q}-z_{q'})(z_{q}-z_{l})}\right )K_{qq'}
(K_{ql}-K_{q'l})  \\ 
\left [ z_{q},z_{q'}\right ] &&= 0 \\  
\left [ d_{q},z_{q'}\right ] &&= \delta_{qq'}\left ( 1 + \sum_{l} a_{ql}K_{ql}
\right ) - a_{qq'}K_{qq'}
\end{eqnarray}
so the $W_{\infty}$ algebra structure is now lost. In the order to see if and 
how this symmetry can be recovered we note that the first two $W_{\infty}$ 
generators are:
\begin{eqnarray}
{\cal W}^{1}_{m}(d) &&= \sum_{q}z_{q}^{m+1}\frac{\partial}{\partial z_{q}} + 
\sum_{l>l'}a_{ll'}\frac{z_{l}^{m+1}-z_{l'}^{m+1}}{z_{l}-z_{l'}}(1-K_{ll'})
\\ 
W_m^2(d) &&=\sum_{q} z_{q}^{m+2}\frac{\partial^{2}}{\partial z_{q}^{2}} 
+ 2 \sum_{l>l'}a_{ll'} 
\frac{z_{l}^{m+2}\frac{\partial}{\partial z_{l}}- 
z_{l'}^{m+2}\frac{\partial}{\partial z_{l'}}}{z_{l}-z_{l'}} 
\\ &&- 
\sum_{l>l'}a_{ll'}\frac{z_{l}^{m+2}+z_{l'}^{m+2}}{(z_{l}-z_{l'})^{2}}
(1-K_{ll'}) \nonumber \\
&&-
\sum_{q,l\neq l'\neq q}\frac{a_{lq}a_{ql'}z_{q}^{m+2}}{
(z_{l}-z_{q})(z_{q}-z_{l'})}(1-K_{lq})(1-K_{ql'}) \nonumber 
\end{eqnarray}
where we have used 
\begin{eqnarray}
d^{2}_{q} &&=\frac{\partial^2}{\partial z^{2}_{q}} +
\sum_{l\neq q}\frac{a_{ql}}{z_{q}-z_{l}}\left [
\frac{\partial_{q}}{\partial z_{q}} - \frac{\partial_{l}}{\partial z_{l}}+   
(\frac{\partial_{q}}{\partial z_{q}}+\frac{\partial_{l}}{\partial z_{l}})
(1-K_{ql})\right ] \\  &&- 
\sum_{l\neq q}\frac{a_{ql}}{(z_{q}-z_{l})^{2}}(1 - K_{ql}) +
 \sum_{q,l\neq l'\neq q}\frac{a_{ql}a_{ql'}}{(z_{q}-z_{l})(z_{q}-z_{l'})}
(1 - K_{lq})(1 - K_{ql'}) \nonumber 
\end{eqnarray}

If all $a_{ll'}=r^{2}$ they still close a $W_{\infty}$ algebra on the states 
that we have used to 
realize them on which the exchange operators $K_{qq'}$ becomes a c-number
\begin{equation}
\left [W^{n}_{m}(d) ,W^{n'}_{m'}(d) \right ]\phi (z_{q}) =  
\left [W^{n}_{m} ,W^{n'}_{m'} \right ]\phi (z_{q})
\end{equation}
where $W^{n}_{m}$ are the $W^{n}_{m}(d)$ with $K_{ll'}=1$.

In particular $W_0^2(d)$ becomes the effective CS hamiltonian when it is 
applied on the symmetric functions.

If we use the freedom of adding any antisymmetric term that cannot give 
observable results, we restore the $W_{\infty}$ structure that is realized only 
modulo the transformations generated by the operators of Sec.(4.2).

If we define the operator $L_{q}$ as
\begin{eqnarray}
L_{q} &&= {\cal F}^{\{0,0,0,b_{ll'}\}}_{N}(z_{i_{l}},z_{j_{l'}}) d_{q} 
{\cal F}^{\{0,0,0,b_{ll'}\}-1}_{N}(z_{i_{l}},z_{j_{l'}}) \\  
&&= \frac{\partial}{\partial z_{q}} -
\sum_{l\neq q}\frac{1}{z_{q}-z_{l}}[b_{ql}-a_{ql}(1 -\eta (b_{ql})K_{ql})] 
\nonumber 
\end{eqnarray}
the $ L_q $ satisfy the same algebra than the Dunkl
operator $d_{q}$ with $K_{ll'}\rightarrow \eta_{ll'}(b_{ll'})K_{ll'}$.

Now we have
\begin{eqnarray}
{\cal W}^{1}_{m} &&= {\cal F}^{\{0,0,0,b_{ll'}\}}_{N}(z_{i_{l}},z_{j_{l'}})
W_m^1(d) {\cal F}^{\{0,0,0,b_{ll'}\}-1}_{N}(z_{i_{l}},z_{j_{l'}}) \\ &&= 
\sum_{q}z_{q}^{m+1}\frac{\partial}{\partial z_{q}} - 
\sum_{l>l'}\frac{z_{l}^{m+1}-z_{l'}^{m+1}}{z_{l}-z_{l'}}
[ b_{ll'}-a_{ll'}(1 - \eta (b_{ll'})K_{ll'})] 
\end{eqnarray}
and
\begin{eqnarray}
{\cal W}_m^2 &&={\cal F}^{\{0,0,0,b_{ll'}\}}_{N}(z_{i_{l}},z_{j_{l'}})
W_m^2(d) {\cal F}^{\{0,0,0,b_{ll'}\}-1}_{N}(z_{i_{l}},z_{j_{l'}}) \\ &&=
\sum_{q} z_{q}^{m+2}\frac{\partial^{2}}{\partial z_{q^{2}}} 
- 2\sum_{l>l'}(b_{ll'}-a_{ll'})
\frac{z_{l}^{m+2}\frac{\partial}{\partial z_{l}}- z_{l'}^{m+2}
\frac{\partial}{\partial z_{l'}}}{z_{l}-z_{l'}} 
\\ &&+
\sum_{l>l'}\frac{z_{l}^{m+2}+z_{l'}^{m+2}}
{(z_{l}-z_{l'})^{2}}
[ (b_{ll'}-a_{ll'})b_{ll'}-b_{ll'}(a_{ll'} -1)-a_{ll'}(1 - \eta (b_{ll'})
K_{ll'})] \nonumber \\ 
&&+ \sum_{q,l\neq l'\neq q}\frac{z_{q}^{m+2}}{(z_{l}-z_{q})(z_{q}-z_{l'})} 
\times[b_{lq}(b_{ql'}-a_{ql'})+b_{ql'}(b_{lq}-a_{lq}) \nonumber \\ 
&&-b_{lq}a_{ql'}-b_{ql'}a_{lq}-
a_{lq}a_{ql'}(1-\eta (b_{ql})K_{lq})(1-\eta (b_{ql'})K_{ql'})] \nonumber 
\end{eqnarray}

When $a_{ll'}=a$ and $b_{ll'}=b$ it is possible to prove on the line of 
the proof of Ref. \cite{37} that the
commutators of any symmetrized operators do not contain any extra terms
depending on $K_{ll'}$, i.e. the abstract algebra does not depends on the
statistics of the particles.
Moreover by a translation of the q-th Dunkl operator by $z_q$ we can introduce
an harmonic oscillator potential in the hamiltonian.

Notice that if we consider the case $ b_{ll'}=a_{ll'}$ the generator 
${\cal W}_0^2$ becomes the multispecies hamiltonian of CS model with a 
three-body potential plus terms that contain the projectors 
$(1-\eta (b_{ql})K_{lq})(1-\eta (b_{ql'})K_{ql'})$ vanishing on the relative 
wave functions.

Now we are in position to discuss the CS model with spin. 
We introduce an internal space (spin space of dimension $2s+1$) so that the 
true wave function is the tensor
product of the $\Psi (z_{q})$ with a spin wave function $\chi$. 
We can simulate the
action of the spin by introducing a $(2s+1)$-dim ``vacuum"  such that

\begin{equation}
P_{ll'} |\chi > = \eta_{ll'}(s) |\chi > 
\end{equation}

where now we write the exchange operator as the product of an operator
$K_{ll'}$ acting on the labels $ll'$ of the excited states and the operator
$P_{ll'}$ acting on the spin vacuum. 
Now the equation, which the eigenstates of the collective hamiltonian have to 
satisfy, becomes
\begin{equation}
 (P_{ll'} - K_{ll'}) \phi '(z_{q})|\chi > = 0
\end{equation}
where $\phi '(z_{q})$ can be related to the symmetric function $\phi(z_{q})$ by a 
Jastrow like function that gives the same phases of spin vacuum:
\begin{equation}
\phi '(z_{q}) = \phi (z_{q}) {\cal F}^{s}(z_{l},z_{l'})
\end{equation}   

On this space the Dunkl operator is written as
\begin{equation}
d_{q} = \frac{\partial}{\partial z_{q}} - \sum_{l\neq q} 
\frac{a_{ql}}{z_{q}-z_{l}}(P_{ll'} - K_{ll'})
\end{equation}

The above formula is a particular case ($\mu = \nu$) of the general case
considered in \cite{36}. 
The general case can be reproduced in our approach
by a translation of any root $r_q (r_q^2 = \nu)$ by a quantity $t_q$ such
that $\mu = (r_q + t_q)^2$ and by defining the generator  $L_{q}$ with
the Jastrow function which depends only on $\mu - \nu = t_q(t_q + r_q)$,
that is the generators of the differential algebra are written with a factor
which does not take in to account completely the statistics (case 
$b_{ll'}\neq a_{ll'}$). 

It is interesting to look more in detail at this interpretation of the
statistics in terms of shift of the roots. The free case (no interaction)
is obtained when $r^{2} = 0$, i.e. when $r$ is a light-like vector. So
 the lattice $\Lambda$ can be considered in a Minkowskian space. 

If we take $\lambda$ in the dual lattice:
\begin{equation}
\lambda = \alpha + n_{+}K_{+} + n_{-}K_{-} \;\;\;\;\; n \in Z_{+}
\end{equation}

where $K_{+}, K_{-}$ are the two lightlike vector such that 
$K_{+} \cdot K_{-} = 1$ and $\alpha$ is a vector orthogonal to $K_{+}$ and
$K_{-}$. 

The interaction with statistics $\nu$ can be introduced by translating
the lightlike roots (let us say $r_q = r = nK_{+}$ for any q) by $t$ a vector
which requiring the invariance of the total momentum ($ \sum_{q} r_q \cdot
\lambda$) can be written as:
\begin{equation}
t\cdot \lambda = 0 
\end{equation}
The introduction of the spin requires the extension of the algebra of the
observables to a charged $W$-algebra. 

Notice that the ground 
state function has an expression very similar to the measure of the Selberg
correlation integrals, whose relevance for the CS models have been discussed
in \cite{18}. Many relations between CS model and other topics, for instance KdV
equations, can by simply understood in this framework.

It is an interesting point to study the action of the operators of Sec.(3), that 
give  amplitudes of GVOs, in the case of CS model.

We restrict the discussion to the case of VOs and apply the operator to obtain 
the CS ground state of $N-1$ particles from $N$ one:
\begin{eqnarray}
\psi^{\{r'_{q},\lambda\}}_{N-1}(z_{q}) &&= \nabla^{\{r_{N}\cdot r_{N-1}\}}
(z_{N},z_{N-1})\psi^{\{r_{q},\lambda\}}_{N}(z_{q}) \\ 
&&= \lim_{z_{N}\rightarrow z_{N-1}} 
(z_{N}-z_{N-1})^{-r_{N}\cdot r_{N-1}}\psi^{\{r_{q},\lambda\}}_{N}(z_{q}) 
\nonumber 
\end{eqnarray} 
where
\begin{eqnarray}
r'_{q} &&= r_{q} ~~\forall ~ q=1,\dots ,N-2  \nonumber \\ 
r'_{N-1} &&= r_{N}+r_{N-1} 
\end{eqnarray}
We hope that this formulation can be applied on second quantization of CS 
model where $\nabla $ operator can be identified with annihilator operator.
  
\bigskip

\section{Conclusions}

We have shown that by a similarity transformation ("dressing") of the free 
differential operators generating the
$ W_{\infty} $ , we still preserve the algebraic structure and that with
a particular choice of the "dress" we can identify the CS hamiltonian with 
the generator $ W_0^2 $. 

The choice of ``dressing" depends on two aspect, the first is related to the 
existence of non-unitary similarity transformations that give the $W_{\infty}$ 
algebra for any GVO that is discussed in Sec.(2)-(3).

The second gives a sector independent basis by using the permutations 
operator $K_{ll'}$ that takes in account the multiply connected configuration 
space  for CS model.
In term of vertex algebra this is just the duality invariance property of 
amplitudes.    

The dressing function is given by the correlation 
function of the product of N VOs, computed out of mass shell 
between an arbitrary in state and an out state fixed by the choice of the
in state and by the value of the roots. The $W$ algebra is so related with the
Ward identities for VOs amplitudes, identities always satisfied also
by correlation functions for the product of generic GVOs. 
Although we have not yet explicitly computed the correlation function
for product of any GVO, we have presented here the whole formalism to
emphasize the role, we believe fundamental, of the vertex algebra.
In the case of the product of N VOs with the same root, corresponding to the
CS model, we find all the results of \cite{28}, but our formalism allows
to get more general equations with potential whose coupling constant is
not necessarily equal for all the particles. 
The system of N VO appears as a system of N particles, of equal mass, with 
internal quantum numbers specified by the roots $r_i$. The particles are
identical if their roots are equal. The observables of a system of
identical particles must be symmetric operators, so operators not
belonging to the diagonal $W_{\infty}$, must be not observable.
For instance we have (${\cal L}_{i}$ are the operators given by 
eq:(\ref {eq: 46})) 

\begin{equation}
\left (
z_1{\cal L}_1(z_1) \; - \; z_2{\cal L}_1(z_2) \psi_N(z_{q}) \right )
= (r_1 - r_2) \cdot \lambda \psi_N(z_{q})
\end{equation}  

and the r.h.s. is vanishing if $r_1= r_2$. More generally
as the momenta defined on the circle have to be quantized, the product
$ r_q \cdot \lambda$ must be an integer, that implies that  $\lambda$ 
has to belong to the dual lattice of the lattice $\Lambda$ of the roots.
So translating $\lambda$ by an element of the lattice does not change the
value of the relative momenta (up to an integer), so the translation of
$\lambda$ can be interpreted as a Galilean boost of the system.
So the algebra of observables is the whole algebra
of differential operators 
One may expect that a classification of all the possible vertex operators 
may give a classification of all the possible integrable models.  
Let us emphasize once more that we have shown that the invariance for
a $ W_{\infty} $ algebra, which in the literature has been established
only for particular models, is a general feature, consequence of the
algebra of differential operators for correlation functions.
Moreover the connection between integrable 1+1 models as CS and  vertex 
algebra is given by the intrinsic structure  of GVOs that are an explicit 
realization of the vertex algebra. 

We conjecture that ``integrability conditions" as Yang-Baxter equation, can be 
deduced by properties of vertex algebra as it happens in the simplest case of CS 
model.

A very interesting question to understand is if there are physical models
corresponding to correlation functions of the product of arbitrary GVOs.
Independently of any physical interpretation or interest it is natural to
raise the question if
such models are integrable. We believe that the answer is negative, at least
if integrability is meant in the usual sense. In fact integrability
requires that $r^{2}$ in eq.(56) be non negative. This condition is not
guaranteed either for the roots or for the weight for a generic (not affine) 
Kac-Moody algebra.

It is interesting to discuss the connections and the differences between the 
structure of the integrable non relativistic models and the structure of 
string theory, where the Lorentzian algebras play an essential role. 
For instance many of the above considerations still hold replacing the
term particle with the term string.
However a thorough discussion of this topic requires further analysis and 
it will be eventually carried out elsewhere.

Finally let us remark that $W_{\infty}$ algebras are related to area preserving 
diffeomorphisms and these structures arise as a property of vertex algebra 
independently of any physical models.

\bigskip

\pagebreak

\pagebreak

\begin{center}
{\bf APPENDIX }
\end{center}

\bigskip 

To give an explicit expression of amplitudes and GVOs products we need to 
compute any arbitrary $\lmschik\ $ function. This can be done in the following 
way.

By definition:
\begin{equation}
\lmschik\ =\lim_{\{z_{v_{q}}\rightarrow z_{q}\}} \mschi\ 
\end{equation}
and 
\begin{equation}
\mschi\ = \left ( \partial^{\{k_{v_{q}}\}}(z_{v_{q}})\mgcaf\ \right ) 
\end{equation}

Factorizing ${\cal F}$ in terms of $q$-variable:
\begin{eqnarray}
&& \mgcaf\ = \prod_{l>l':l,l'\neq q}\cafl\ \\    
&&\times \prod_{l>q} {\cal F}^{\{a_{i_{l}v_{q}},a_{i_{l}q},a_{lv_{q}},a_{lq}\}}_{lq}
(z_{i_{l}},z_{v_{q}})
\prod_{l'<q} {\cal F}^{\{a_{v_{q}j_{l'}},a_{v_{q}l'},a_{qj_{l'}},a_{ql'}\}}_{ql'}
(z_{v_{q}},z_{j_{l'}}) \nonumber
\end{eqnarray}

we indicate  with $k_{v_{q}}^{l}$ and $k_{v_{q}}^{l'}$ the number of 
derivatives that act, respectively, on terms 
${\cal F}_{lq}$ and  ${\cal F}_{ql'}$, these numbers must satisfy the 
identity:
\begin{equation}
\sum_{l>q} k_{v_{q}}^{l}+\sum_{l'<q} k_{v_{q}}^{l'}=k_{v_{q}}~~~
\forall ~\{q, v_{q}\} \label {eq: LMSCHIK}
\end{equation} 

At this point it is possible to give a factorization of $\mschi\ $ in terms 
of two-point functions:
\begin{equation}
\mschi\ =
\sum_{\{\sum_{l>q} k_{v_{q}}^{l}+\sum_{l'<q} k_{v_{q}}^{l'}=k_{v_{q}}\}}
\prod_{l>l'}
{\cal S}^{\{a_{i_{l}j_{l'}},a_{i_{l}l'},a_{l'j_{l'}},a_{ll'}\}}_{\{k_{i_{l}}^{l'}
k_{j_{l'}}^{l}\}}(z_{l},z_{l'}) \nonumber 
\end{equation}
where indices $l,l'$ include also the renamed $q$, and, by means of 
substitutions:
\begin{eqnarray}
&& r_{i_{l}}\rightarrow r_{i}, ~~~ r_{l}\rightarrow r, ~~~ 
k_{i_{l}}^{l'}\rightarrow  k_{i} ~~~\forall ~l,l' \\ 
&& r_{j_{l'}}\rightarrow s_{j},~~~ r_{l'}\rightarrow s, ~~~ 
 k_{j_{l'}}^{l}\rightarrow k_{j} ~~~\forall ~l,l' \\ 
\end{eqnarray}

a general ${\cal S}$ is:
\begin{equation}
\schi\ =\left ( \partial^{\{k_{i},k_{j}\}}(z_{i},\xi_{j}) \gcaf\ \right )  
\end{equation} 

In this way it is necessary to compute only this class of functions.

By the factorization of $\gcaf\ $ it is possible give an explicit expression 
for its derivatives:
\begin{eqnarray}
&& \partial^{k_{i}}(z_{i})\partial^{k_{j}}(\xi_{j})
\gcaf\ =
\sum_{\{\sum_{i}p_{i}=k_{i}, \sum_{j}p_{j}=k_{j}\}}
\sum_{\{l_{i},l_{j}\}}^{\{p_{i},p_{j}\}} \\ 
&& \times\prod_{i,j}
\left [ \partial^{l_{i}}(z_{i})\partial^{l_{j}}(\xi_{j})
(z_{i}-\xi_{j})^{a_{ij}}\right ]
\left [ \partial^{p_{i}-l_{i}}(z_{i})(z_{i}-\xi)^{a_{i0}}\right ]
\left [ \partial^{p_{j}-l_{j}}(\xi_{j})(z-\xi_{j})^{a_{0j}}\right ] \nonumber 
\end{eqnarray}

that can be written in a more compact form introducing the coefficients: 
\begin{equation}
\chia\ = (-1)^{\sum_{j}l_{j}}\prod_{i,j}
\left (\begin{array}{cc} a_{i0} \\p_{i}-l_{i}\end{array}\right )
\left (\begin{array}{cc} a_{ij} \\l_{i}+l_{j}\end{array}\right )
\left (\begin{array}{cc} a_{0j}\\p_{j}-l_{j}\end{array}\right )  
\end{equation}

\begin{eqnarray}
\schi\ &&=\gcaf\ \\ &&\times  
\sum_{\{\sum_{i}p_{i}=k_{i}, \sum_{j}p_{j}=k_{j}\}}
\sum_{\{l_{i},l_{j}\}}^{\{p_{i},p_{j}\}}
\frac{\chia\ }{ 
(z_{i}-\xi_{j})^{l_{i}+l_{j}}(z_{i}-\xi)^{l_{i}}(z-\xi_{j})^{l_{j}}} \nonumber 
\end{eqnarray}

To give a final formula for $\lmschik\ $ it is necessary to take the limits 
the $z_{i}\rightarrow z$ and $\xi_{j}\rightarrow \xi $ for all $i,j$:
\LSCHI\ 

where 
\CHIS\ 

Replacing this formula in the eq:(\ref {eq: LMSCHIK}) with the obvious 
renaming of indices:
\begin{eqnarray}
&& r_{i}\rightarrow r_{i_{l}}, ~~~ r\rightarrow r_{l}, ~~~ 
k_{i}\rightarrow  k_{i_{l}}^{l'} ~~~\forall ~l,l' \\ 
&& s_{j}\rightarrow r_{j_{l'}},~~~ s\rightarrow r_{l'}, ~~~ 
k_{j}\rightarrow  k_{j_{l'}}^{l} ~~~\forall ~l,l' 
\end{eqnarray}

we obtain the final formula:
\begin{equation}
\lmschik\ = 
\sum_{\{\sum_{l>q} k_{v_{q}}^{l}+\sum_{l'<q} k_{v_{q}}^{l'}=k_{v_{q}}\}}
\prod_{l>l'}
\frac{\chil\ }
{(z_{l}-z_{l'})^{-r_{l}\cdot r_{l'}+\sum_{i_{l}}k_{i_{l}}^{l'}
+\sum_{j_{l'}}k_{j_{l'}}^{l}}} 
\end{equation}

Q.E.D.

\pagebreak


\begin{thebibliography}{99}


\bibitem{14} F. Calogero - J. Math. Phys. \underline{12}, 419 (1971)

\bibitem{15} B. Sutherland - Phys. Rev. A\underline{5}, 1372 (1972)

\bibitem{22a} K. Hikami and M. Wadati - J. Phys. Soc. Jap. \underline{62},
             3857 (1993)

\bibitem{24} Satoshi Iso - Nucl. Phys. B \underline{443}, 581 (1995) 

\bibitem{16}  M.A. Olshanetsky and A.M. Perelomov - Phys. Rep. \underline{94}, 313 (1983)
 
\bibitem{17} P.I. Etingof - J. Math. Phys. \underline{36}, 2636 (1995)

\bibitem{21} H. Ujino, M. Wadati and K. Hikami - J. Phys. Soc. Jap. \underline{62},
             3035 (1993)

\bibitem{23} D. Bernard, K. Hikami and M. Wadati - Preprint SPhT-94-155

\bibitem{22b} K. Hikami and M. Wadati - J. Phys. Soc. Jap. \underline{62},
             4203 (1993)

\bibitem{22c} A.P. Polychronakos - Phys. Rev. Lett. \underline{69}, 703 (1992)

\bibitem{22d} L. Brink, T.H. Hansson and M.A. Vasiliev - Phys.Lett.B 
       \underline{286}, 109 (1992)

\bibitem{22e} E. Bergshoeff and M.A. Vasiliev - Int.J.Mod.Phys.\underline{10}, 
         3477 (1995)

\bibitem{19} I.B. Frenkel, J. Lepowsky and A. Meurman - {\it Vertex Operator Algebras
     and Monster}, Pure and Applied Mathematics, Vol. 134, Academic Press,
     San Diego (1988)

\bibitem{5} R.W. Gebert - Int. J. Mod. Phys. A \underline{8}, 5441 (1993)

\bibitem{7} V. Marotta and A. Sciarrino - J. Phys. A \underline{26}, 1161 (1993)

\bibitem{6} V. Marotta and A. Sciarrino - hep-th/9405192 

\bibitem{11} V. Marotta - PHD Tesis - Universit\'a di
Napoli ``Federico II" (1995) (unpublished)

\bibitem{1} V.G. Kac - Funct. Anal. Appl. \underline{1}, 328 (1967)

   R.V. Moody - Bull. Amer. Math. Soc. \underline{73}, 217 (1967)

\bibitem{4} R. Borcherds - J. Algebra \underline{115}, 501 (1988)

\bibitem{25}N.H. Jing and T. Jozenak -  Duke Math. J. \underline{67}, 377 (1992);

            N.H. Jing - J. Algebraic Combinatories \underline{3}, 291 (1994);

            T.H. Baker - J. Phys. A \underline{28}, 589 (1995)

\bibitem{8} P. Goddard and D.I. Olive - {\em Algebras, Lattices and Strings } 
in ``Vertex Operators in Mathematics and Physics",
   ed. J. Lepowski, S. Mandelstam and I. Singer (Springer Verlag, Heidelberg, 
1984)

\bibitem{30} C. Dong and J. Lepowsky - {\em Abelian interwining algebras - a 
generalization of vertex operator algebras}   

\bibitem{27} E. Witten and B. Zwiebach - Nucl. Phys. B\underline{377}, 55 (1992)

\bibitem{33} J. M. Leinaas and J. Myrheim - Phys. Rev. B\underline{37}, 9286 
(1988); Int. J. Phys. 

B\underline{5}, 2573 (1991); Int. J. Mod. Phys.  A\underline{8}, 3649 (1993)  

\bibitem{31} H. Awata, Y. Matsuo, M. Fukima and S. Odake - Prog. Theor. Phys. 
Supp. \underline{118}, 343 (1995) 
            
\bibitem{34} A. Cappelli, C.A. Trugenberger and G.R. Zemba - Phys. Rev. Lett. 
 \underline{72}, 1902 (1994)

\bibitem{18}  P.J. Forrester - Nucl. Phys. B\underline{388}, 671 (1992); 

ibidem  B\underline{416}, 377 (1994)

\bibitem{28} H. Awata, Y. Matsuo, S. Odake and J. Shiraishi - Nucl. Phys. 
             B\underline{449}, 347 (1995) 

\bibitem{37} S.B. Isakov and J. M. Leinaas - HEP-TH/9510184

\bibitem{38} R. Stanley - Adv. Math. \underline{77} 76 (1989)

\bibitem{36} L. Brink and M. A. Vasilev - Mod. Phys. Lett.  A\underline{8} 
             3585 (1993) 

\bibitem{40} F. D. M. Haldane - Phys. Rev. Lett.  \underline{67},
             937 (1991)

\bibitem{41} C. Furtlehner and S. Ouvry- Mod. Phys. Lett.  B\underline{9}, 
             503 (1995)

\bibitem{42} D. Sen - Phys. Rev. Lett.  \underline{67},
             937 (1991)

\end{thebibliography}
\end{document}